\documentstyle[aps,12pt,preprint]{revtex}

\begin{document}

\newcommand{\nc}{\newcommand}
\newcommand{\BE}{\begin{equation}}
\newcommand{\EE}{\end{equation}}
\newcommand{\BA}{\begin{eqnarray}}
\newcommand{\EA}{\end{eqnarray}}
\newcommand{\V}[1]{\mbox{\boldmath $#1$}}

\title{Thermal Photon Emission from QGP Fluid}
\author{Tetsufumi Hirano$^1$  \thanks{Electronic address :
695l0943@cfi.waseda.ac.jp}, Shin Muroya$^2$
\thanks{Electronic address : muroya@yukawa.kyoto-u.ac.jp},
and Mikio Namiki$^1$ \thanks{Electronic address :
namiki@mn.waseda.ac.jp}}

\address{$^1$Department of Physics, Waseda University\\Tokyo
169, Japan}
\address{$^2$ Tokuyama Women's College\\Tokuyama, Yamaguchi
754, Japan}
\date{\today}

\preprint{WU-HEP-96-13,TWC-96-1}
\maketitle
\begin{abstract}

We compare the numerical results of thermal photon
distribution from the hot QCD
matter produced by high energy nuclear collisions,
based on hydrodynamical model, with the recent experimental
data obtained by CERN WA80.
Through the asymptotic value of the slope
parameter of the transverse momentum distribution, we 
discuss the characteristic temperature of the QCD fluid.

\noindent PACS number(s) : 12.38.Mh, 25.75.+r, 24.85.+p

\end{abstract}
\newpage
\section{INTRODUCTION}

One of the most important problems in the recent high energy
physics is to analyze the quark-gluon plasma (QGP) produced in 
ultra-relativistic nuclear reactions \cite{QM95}.

In a previous paper \cite{Mizutani88}, we formulated a
semi-phenomenological quantum transport theory for 
quark-gluon plasma fluid based on an operator-valued
Langevin equation. Giving a mode spectrum and a damping as
the input into this theory, we can easily calculate thermodynamical
quantities and transport coefficients. Along this line of
thought, we have already
discussed the space-time evolution of the (1+1)-dimensional
viscous quark-gluon plasma fluid with phase transition
\cite{Akase89}, the (3+1)-dimensional perfect fluid quark-gluon
plasma \cite{Akase91} and the baryon-rich
quark-gluon plasma \cite{Ishii92}.
Furthermore we have also analyzed the existing
experiments based on the theoretical results of the
(3+1)-dimensional perfect fluid model with phase transition
\cite{Muroya95}.

In high energy nuclear collisions, all secondary particles,
such as hadrons, 
leptons and photons, come out from a hot and dense matter
which is produced at the earlier stage of the nuclear
collision. We first naively expect that the final
distribution of those particles directly reflects
information of the hot and dense matter, as the particle
source. As for hadrons, however, we can say
that the final distribution is seriously masked
by strong final interactions. For this reason, we prefer to
observe leptons and/or photons directly coming from the
particle source rather
than hadrons themselves, in order to
study the particle source in the earlier state at higher
temperature and at higher density. On the other hand, we
expect to observe a sort of phase transition between the
hadron and quark-gluon plasma states in the course of 
cooling down of the matter from higher to low
temperature. For this purpose,
the lepton distribution would be the best to be observed,
even though the observation must not be so easy from the
practical point of view.
Instead, if we can distinguish experimentally
photons which are thermally emitted from the particle source
at higher temperature and higher density,
from photons emerging through the process $\pi^0 \rightarrow 2 \gamma$,
the observation of these thermal photons would be meaningful.

First imagine that we have a simple case in which the
average wavelength of
photons be much longer than the size of the matter, even
though this case is
very far from realistic cases. In this case we can expect to have 
a simple dipole radiation emitted from an oscillating dipole along 
the collision axis, and then to observe photons mainly
distributing around
the transverse direction. The distribution is to be compared
with photons 
coming from the decay of neutral pions, which sharply
distribute around the
collision axis. This expectation must not be true, because
the above assumption that the wavelength is much longer than
the particle source is not justified. 
However, we could expect to observe a certain trend of
expansion in the $p_{T}$-distribution of high energy photons
produced in high energy nuclear collisions.
In this paper, we examine this kind of trend by making use
of numerical simulations of hydrodynamical expansion of high
temperature and high density quark-gluon plasma fluid.

Since the thermal photon is considered to keep the information
about the early stage of the hot matter produced by relativistic
nuclear collisions, many theoretical analyses have already been
done.
Some groups \cite{Srivastava,Arbex,Neumann,Dumitru} have
analyzed the experimental data of CERN WA80 S+Au
200GeV/nucleon (preliminary) \cite{94WA80,95WA80} so as 
to fit their theoretical model to the thermal photon
emission data.
Most of these papers except Ref. \cite{Dumitru}, however,
dealt only with the photon spectrum leaving  
the hadron spectrum not analyzed.
In this paper we analyze the photon and the hadron
distribution produced by the hot QCD matter
in a consistent way.

1) We first choose parameters in the hydrodynamical
model so as to reproduce the hadronic spectrum, {\it i.e.}
the (pseudo-)rapidity distribution and the transverse momentum
distribution.

2) We derive the thermal production rate of photons from a
unit space-time volume based on the finite temperature 
field theory.

3) Accumulating the thermal production rate over the whole
space-time region covered with the particle source, which is 
estimated by the hydrodynamical model,
we evaluate the thermal photon distribution which is to
be compared with the experimental data.

Assuming local equilibrium for the hot QCD matter 
which will be produced in high energy nuclear collisions, 
we will apply the 
hydrodynamical models \cite{Akase91} \cite{Muroya95}.
As for the equation of state, we discuss two different types 
of the model,
{\it i.e.}, the QGP fluid model with phase transition
between QGP and hadrons and the hot hadron gas model without 
phase transition.
Based on the assumption that the QGP fluid of the dominant
mode obeys the operator-valued Langevin equation
\cite{Mizutani88}, we can easily deal with the thermal
photon and obtain the formula for the production rate.
By using the formula, we derive the thermal photon distribution 
in high-energy nuclear reactions, and compare the
theoretical results given by the QGP fluid model with phase
transition with the hot hadron gas model without phase
transition in detail.

In Sec.\,I$\!$I, we shortly review the relativistic hydrodynamical
model with phase transition.
In Sec.\,I$\!$I$\!$I, using the quantum Langevin
equation, we derive the production rate of thermal photons
from QCD matter at the high temperature region produced by high
energy nuclear reactions.
The transverse photon distribution and the 
asymptotic slope parameter will be obtained in Sec.\,I$\!$V.  
Section V is devoted to concluding remarks.

We use the natural unit ($\hbar = c = 1$ together with $k_B 
= 1$) throughout this paper.

\section{HYDRODYNAMICAL MODEL WITH PHASE TRANSITION 
AND PARTICLE PRODUCTION}

The hydrodynamical equation for perfect fluid is given by
\BA
\partial_\mu T^{\mu\nu} & = &0, \\
T^{\mu\nu} & = & (E+P)U^\mu U^\nu - P g^{\mu\nu},
\EA
where $E$, $P$, and $U^\mu$ are,
respectively, energy density, pressure, and local four
velocity.
Energy density, pressure, and entropy density are given by
\BA
E(T) &=& \frac{1}{2\pi^2}\int_{0}^{\infty}p^2 dp \varepsilon
(\V{p}) n(\varepsilon(\V{p}),T),\\
P(T) &=& \frac{1}{6\pi ^2}\int_{0}^{\infty}p^2 dp
\frac{p^2}{\varepsilon(\V{p})} n(\varepsilon(\V{p}),T),\\
S(T) &=& \frac{E+P}{T}.
\EA

Following Ref. \cite{Akase91}, we assume a simple model for the mode
spectrum of the fluid
\BE
\varepsilon (\V{p}) = A \sqrt{\V{p}^2+M^2}
\frac{1-\tanh{\frac{T-T_C}{d}}}{2}+\mid \V{p} \mid
\frac{1+\tanh{\frac{T-T_C}{d}}}{2}.
\EE
Here we suppose that the fluid in the QGP phase is
dominantly composed by u-, d-,
s-quarks and gluons and that the fluid in the hadron phase
is dominantly composed  
by pions and kaons. In this case we put $A$ $=$ 1.89,
$M$ $=$ 200 MeV, $T_c$ $=$ 160 MeV, and $d$ $=$ 2 MeV based
on a previous analysis.
With these parameters, we obtain the phase
transition-like behavior of energy density (see Fig.\,1), which
seems to reproduce the Lattice QCD result \cite{Akase89}.

In a previous paper \cite{Muroya95}, by making use of the
following two models:
1) the QGP fluid model with phase transition,
2) the hot hadron gas model without phase transition,
we have analyzed the pseudo-rapidity distribution
of charged hadrons in S+Au 200 GeV/nucleon collision
obtained by CERN WA80 \cite{92WA80}.
In this paper in order to develop the hydrodynamical model
furthermore,
we are going to analyze the $p_T$-distribution of neutral
pions also given
by CERN WA80 \cite{94WA80}
as well as the pseudo-rapidity distribution of charged 
hadrons.
According to the previous analysis \cite{Muroya95}, 
we use the first model (the QGP fluid model with phase
transition) specified by the
initial temperature
$T_i$ = 195 MeV, the critical temperature $T_c$ = 160 MeV,
and the freeze-out temperature $T_f$ = 140 MeV, and the
second model (the hot hadron gas model without phase transition)
specified by $T_i$ = 400 MeV and $T_f$ = 140 MeV.
We choose other parameters being the same as given by
Ref.\cite{Muroya95}.
For these models, we obtain theoretical
results of the hadronic spectrum.
See Fig.\,2 and Fig.\,3.
From Fig.\,2 and Fig.\,3, 
we observe that the both models can consistently reproduce the
experimental data. The hot hadron gas model should have the
initial energy of the fluid exceeding the
total collision energy \cite{Muroya95}, so that we conclude
that the hadron gas model is not accepted.
We discuss the thermal photon distribution based on these two
models in Section I$\!$V, in which 
we will see that the hot hadron gas model
will fail again in the analyses of the photon experimental
data.

\section{THERMAL PRODUCTION RATE OF PHOTONS}

Transition amplitude for a process involving emission of a photon with
momentum $k$ and polarization $\varepsilon^{(\lambda)}$ from
the QCD matter is given by
\BE \label{tra-amp}
T_{\beta\alpha}=\langle
\beta;k,\varepsilon^{(\lambda)};\mbox{out}\mid
\alpha;0;\mbox{in} \rangle,\EE
where $\mid \alpha;0;\mbox{in}\rangle$ and $\mid
\beta;k,\varepsilon^{(\lambda)};\mbox{out}\rangle$ stand for, 
respectively, the initial state and the final state.  
Usual technique of the reduction formula enables us to give
the following transition amplitude
\begin{eqnarray} \label{reduction}
T_{\beta\alpha} & = & \langle \beta;\mbox{out} \mid
a^{(\lambda)}_{out}(k)\mid \alpha;\mbox{in} \rangle \nonumber \\
 & = &
ig^{\lambda\lambda^\prime}\varepsilon^{(\lambda^\prime)}_\mu
\int \frac{d^4 x}{\sqrt{(2\pi)^3 2k_o}} e^{ikx}\langle
\beta;\mbox{out} \mid j^\mu_H (x)\mid
\alpha;\mbox{in} \rangle \nonumber \\
 & = &
ig^{\lambda\lambda^\prime}\varepsilon^{(\lambda^\prime)}_\mu
\int \frac{d^4 x}{\sqrt{(2\pi)^3 2k_o}} e^{ikx}\langle
\beta;\mbox{in} \mid T(Sj^\mu_I (x))\mid \alpha;\mbox{in} \rangle,
\end{eqnarray}
from which we obtain transition probability per unit
space-time volume 
 \begin{eqnarray} \label{R_k_eps1}
R(k,\varepsilon^{(\lambda)})
 & = & \frac{1}{VT}\mid T_{\beta\alpha} \mid^2 \nonumber \\
 & = & \frac{\varepsilon^{(\lambda)}_\mu
\varepsilon^{(\lambda)}_\nu}{VT} \int \frac{d^4x d^4y}{(2
\pi)^3 2k_0}
e^{-ik(x-y)} \nonumber \\
 & \times & \langle\alpha;in \mid \tilde{T}(S^\dagger
j^\mu_I (x))\mid\beta ; in\rangle\langle\beta ; in \mid T(S
j^\nu_I (y)) \mid \alpha; in\rangle.
\end{eqnarray}
Since we are interested in one photon state, we should sum
up with respect to the final states of QCD matter denoted by
$\beta$.
To the lowest order in QED, we obtain
\begin{equation} \label{R_k_eps2}
R(k,\varepsilon^{(\lambda)})=\frac{\varepsilon^{(\lambda)}_\mu
\varepsilon^{(\lambda)}_\nu}{VT} \int \frac{d^4x
d^4y}{(2\pi)^3 2k_0} e^{-ik(x-y)}\langle\alpha;in \mid j^\mu (x)
j^\nu (y) \mid \alpha; in\rangle.
\end{equation}
 
Assuming that a certain mode is dominantly excited in a
local equilibrium system of the hot QCD matter and
the canonical operator of that mode obeys the
{\it quantum} Langevin equation \cite{Mizutani88} (see
Appendix), then we can replace the above matrix element 
with the ensemble average in the sense of quantum Langevin 
equation, {\it i.e.},
\BE
\langle\alpha;in \mid j^\mu(x)j^\nu(y) \mid \alpha; in\rangle
\Longrightarrow \langle j^\mu_{Q.L.} (x)j^\nu_{Q.L.} (y)
\rangle _{Q.L.},
\EE
where subscript Q.L. represents the ensemble average.
Through this paper, we will omit the subscript Q.L. 
for the simplicity.

In the case of bosonic mode (pion or kaon) in the hadron
phase, the source function is given by
\BE \label{j-bos}
j^\mu (x)= i:\phi
^{\dag}(x)(\stackrel{\rightarrow}{\partial^\mu}
-\stackrel{\leftarrow}{\partial^\mu} )\phi (x):.
\EE
The boson field operator in the theory of the quantum Langevin
equation is represented as
\begin{eqnarray} \label{b-field}
\phi (x) & = & \int \frac{d^3 \V{p}}{\sqrt{(2\pi )^3 2\Omega
(\V{p})}}\int_{0}^{\infty} d\omega \sqrt{\frac{\gamma
(T)}{2\pi }}\rho (\V{p},\omega) \nonumber \\
& & \quad \quad \quad \times \left(\frac {A(\V{p},\omega)}
{\omega-E(\V{p},T)} e^{-ipx}+\frac{B^{\dag}(\V{p},\omega)}
{\omega-E^* (\V{p},T)} e^{ipx}\right),\\
E(\V{p},T)& = &\varepsilon (\V{p})-\frac{i}{2} \gamma (T), 
\end{eqnarray}
where $\varepsilon (\V{p})$ = $\sqrt{\V{p}^2+M^2}$ and $\gamma
(T)$ = $cT$ are,
respectively, the mode spectrum and the damping, which we
have given to the theory as input. The free
parameter $c$ in $\gamma (T)$ is so chosen as to fit the
photon transverse momentum distribution to the experimental
data: we have determined $c = 0.01$. 
In the case of fermionic mode (quark) in the QGP
phase, a source function is given by
\BE \label{j-fer}
j^\mu (x)= :\bar{\psi} (x)\gamma ^\mu \psi (x):,
\EE
where
\begin{eqnarray} \label{f-field}
\psi (x)&=&\int \frac{d^3 \V{p}}{\sqrt{(2\pi )^3}}
\int_{0}^{\infty} d\omega \sqrt{\frac{\gamma (T)}{2\pi
}}\rho (\V{p},\omega) \nonumber \\ 
& &\quad \quad \times \sum_{r=1,2} \left( u_r (\V{p},\omega)
\frac{A_r (\V{p},\omega)}{\omega-E(\V{p},T)}
e^{-ipx}+v_r (\V{p},\omega) \frac{B_{r}^{\dag} (\V{p},\omega)}
{\omega-E^* (\V{p},T)} e^{ipx}\right).
\end{eqnarray}
Here we have put $\varepsilon (\V{p})$ = $\mid \V{p} \mid$
and $\gamma (T)$ = $cT$, with $c = 0.01$.
  
Summing up with respect to polarization and wave number, we
can obtain the production rate
\BE
R(T) = \int d^3 \V{k}\sum_\lambda R(k,\varepsilon^{(\lambda)}),\EE
which means the number of the
emitted photons per unit space-time volume of the hot QCD matter.
In order to evaluate the production rate at temperature $T$, 
let us introduce the following simple model
\BE \label{R-PT}
R(T)=R_{had}\frac{1-\tanh\frac{T-T_c}{d}}{2}+R_{QGP}
\frac{1+\tanh\frac{T-T_c}{d}}{2},
\EE
where we have supposed that the phase transition takes place 
around critical temperature $T_c$ within width $d$.
In the infinite $T_c$ limit, we can only keep the hot hadron 
gas model. Based on the Langevin technique mentioned in
Appendix, we can easily calculate $R_{had}$ and $R_{QGP}$ as
\begin{eqnarray} \label{R_had}
&&R_{had}=\sum_s e_s^2 \int \frac{d^3 \V{k}}{(2\pi )^3
2k_0}\int \frac{d^3 \V{p}_1}{(2\pi )^4}\int_{0}^{\infty}
\frac{d\omega_1}{2\omega_1}\frac{\gamma (T)}{\mid \omega_1
-E(\V{p}_1,T)\mid^2} \nonumber \\
&&\qquad \qquad \qquad \qquad \qquad \times \int \frac{d^3
\V{p}_2}{(2\pi )^4}\int_{0}^{\infty}
\frac{d\omega_2}{2\omega_2}\frac{\gamma (T)}{\mid \omega_2
-E(\V{p}_2,T)\mid^2} \nonumber \\
&&\qquad \times (2\pi )^4(-n_b(\omega_1)\{ 1+n_b(\omega_2)\}
(p_1+p_2)^2 \delta^4(p_1-p_2-k)  \nonumber \\
&&\qquad \qquad \quad -\{ 1+n_b(\omega_1)\}
n_b(\omega_2)(p_1+p_2)^2 \delta^4(p_1-p_2+k) \nonumber \\
&&\qquad \qquad \quad -\quad n_b(\omega_1)\quad
n_b(\omega_2)\quad (p_1-p_2)^2 \delta^4(p_1+p_2-k)),
\end{eqnarray}
\begin{eqnarray} \label{R_qgp}
&&\qquad R_{QGP}=\sum_s 3e_s^2 \int \frac{d^3 \V{k}}{(2\pi
)^3 2k_0}\int \frac{d^3 \V{p}_1}{(2\pi )^4}\int_{0}^{\infty}
\frac{d\omega_1}{2\omega_1}\frac{\gamma (T)}{\mid \omega_1
-E(\V{p}_1,T)\mid^2} \nonumber \\
&& \qquad \qquad  \qquad \qquad \qquad \qquad \quad \times
\int \frac{d^3 \V{p}_2}{(2\pi )^4}\int_{0}^{\infty}
\frac{d\omega_2}{2\omega_2}\frac{\gamma (T)}{\mid \omega_2
-E(\V{p}_2,T)\mid^2}\quad \quad  \nonumber \\
&&\quad \quad \times (2\pi )^4(n_f(\omega_1)\{ 1-n_f(\omega_2)\}
(8p_1 p_2-16\sqrt{p_1^2}\sqrt{p_2^2}) \delta^4(p_1-p_2-k)
\nonumber \\
&& \qquad \qquad +\{ 1-n_f(\omega_1)\} n_f(\omega_2)(8p_1
p_2-16\sqrt{p_1^2}\sqrt{p_2^2}) \delta^4(p_1-p_2+k)
\nonumber \\
&& \qquad \qquad +\quad n_f(\omega_1)\quad
n_f(\omega_2)\quad (8p_1 p_2+16\sqrt{p_1^2}\sqrt{p_2^2})
\delta^4(p_1+p_2-k)).
\end{eqnarray}
Here the summation $s$ is over the source particle, and
$e_s$ stands for the electric charge of
a source particle.
These formulas contain three kinds of photon emission
processes, as shown in Fig.\,4.
The double line stands for off-shell particles in thermal
bath.
Taking into account the structure of $R_{had}$ and $R_{QGP}$ 
having factors $\gamma (T)/ \mid \omega-E(\V{p},T)\mid^2$, we 
can understand that the particle spectrum in heat bath is
spread due to the presence of random force.
The production rate as a function of temperature $T$ is
shown in Fig.\,5.

In order to obtain the theoretical formula for
hydrodynamical quantities and transport-theoretical
coefficient, we have assumed that the local system is in
uniform thermal-equilibrium on a microscopic space-time
scale. On the other hand, the QGP fluid hydrodynamically
evolves in the macroscopic space-time region, specified by
space-time coordinate $x$. Solving the hydrodynamical
equation, we obtain the $x$-dependence temperature $T(x)$
and local four velocity $U^\mu(x)$.
Following the well-known procedure, 
we first relate the production
rate in each local system of the QGP fluid with one in the 
whole center-of-mass system of the QGP fluid.
Since the number of photons emitted from the local system
is Lorentz-invariant, we obtain the following relation between
the local system at space-time point $x$ and the overall
center-of-mass system
\BE 
\left.
k_0 \frac{d^3 R_{c.m.}}{d \V{k}^3}=k_0 ^\prime \frac{d^3
R(T(x))}{d \V{k}^{\prime 3}}\right|  _{k_0
^\prime= U^\mu(x) k_\mu}.
\EE
Integrating the above $R_{c.m.}$ over the whole space-time volume 
in which the particle source exists, we obtain momentum
and transverse momentum distributions
\BE
\left.
\frac{d^3 N}{d \V{k}^3} = \int d^4 x
\frac{k_0 ^\prime}{k_0}\frac{d^3 R(T(x))}{d \V{k}^{\prime 3}}
\right| _{k_0 ^\prime = U^\mu(x)k_\mu},
\EE
\BE \label{kt}
\frac{1}{k_T}\frac{dN}{dk_T}=\int d\phi dk_L \frac{d^3 N}{d
\V{k}^3}(k_T,k_L,\phi),
\EE
which are to be compared with experimental data.
Here temperature $T(x)$ and local four velocity $U^\mu(x)$ 
at space-time point $x$ are
given by the numerical solution of the hydrodynamical model.

\section{RESULTS AND DISCUSSION}

Figure 6 shows the numerical results of Eq.\,({\ref{kt}) 
compared with the experimental data (S+Au 200 GeV/nucleon collision)
obtained by CERN WA80 \cite{96WA80}.
The solid curve and the dotted curve are, respectively, 
the contribution
of the hadron phase region and that of the QGP phase region
in the QGP fluid model. 
The whole thermal photon distribution given by our QGP fluid
(with phase transition) model
is the sum of them, but we can easily see that the
contribution of the QGP phase region is negligibly small. 
Using our hydrodynamical model for the QGP fluid with
phase transition, we also see in Fig.\,6 that we reproduce
the experimental data of WA80 well.
The dashed curve stands for the photon distribution given by 
our hot hadron gas model.
The theoretical curve (the dashed curve) deviates from the
experimental data in both absolute value and slope. 
Our formalism keeps a free parameter $c$ in the damping function 
which is directly related to the intensity of
the photon emission but which is almost independent of the slope. 
Even if we choose another value for a free parameter $c$,
the hot hadron gas model cannot reproduce the experimental data
due to the deviation of the slope.
Therefore, we say that \it we reproduce
the WA80 experimental data consistently 
only with the QGP fluid model with phase transition.

\rm
Needless to say, the initial temperature or the phase
transition temperature are very helpful to understand high
energy nuclear reactions, if they are derived from
theoretical analyses as mentioned above.
We naively expect the thermal photon distribution to
directly reflect the characteristic temperature,
but we have to pay attention to the following points:
1) temperature $T$ cannot be invariant under Lorentz
transformations,
as was mentioned in Sec.\,I$\!$I$\!$I,
2) each local system in the QGP fluid has different temperature.
For these reasons, we introduce a little mathematical
manipulation in the following way.
The thermal distribution from
the volume element with velocity $v$ and temperature $T$
should read as
\BA
\left.
\exp(-\frac{k_0}{T})\right| _{\mbox{
\begin{small}
local system
\end{small}
}}
\Longrightarrow & &\left. \exp(-\frac{k_\mu
U^\mu}{T})\right| _{\mbox{
\begin{small}
c.m. system
\end{small}
}}
 \nonumber\\
= & & \exp (- \frac{k_0 U_0 -k_L U_L - k_T U_R \cos \theta}{T}),
\EA
where $ U^\mu = (\frac{1}{\sqrt{1-v^2}},
\frac{\V{v}}{\sqrt{1-v^2}})$.
In order to pick up the most dominant contribution to the
transverse momentum distribution,
we put $k_L= 0$ and rewrite Eq.\,(24) as
\BE
\exp(-\frac{k_T \frac{1-v_T}{\sqrt{1-v^2}}}{T})
= \exp(-\frac{k_T}{\sqrt{1-\frac{v_L^2}{1-v_T^2}}
\sqrt{\frac{1+v_T}{1-v_T}}T}),
\EE
by which we can define the effective temperature $T_{eff} $
of the fluid at the 
volume element with velocity $v_L, v_T$ and temperature $T$
by 
\BE
T_{eff} = 
\sqrt{1-\frac{v_L^2}{1-v_T^2}}\sqrt{\frac{1+v_T}{1-v_T}}T.
\EE
Furthermore we have assumed that the largest value of
$T_{eff}$ dominates in the slope of the transverse momentum
distribution for extremely large $k_T$ region.
The transverse momentum distribution at rapidity $y=0$
and its slope parameters are, respectively, shown in Fig.\,7
and Fig.\,8.
Table 1 shows the maximum $T_{eff}$
given by our numerical results of hydrodynamical simulation
and the slope parameter at $k_T = 20$ GeV for the above two models.
We see in Fig.\,8 that
the slope parameter $T_s$ of each model asymptotically tends 
to maximum value of $T_{eff}$ shown in Table 1.
Through comparison of $T_{eff}$
evaluated by the numerical results of hydrodynamical simulation
with the asymptotic slope parameter $T_s$ of the transverse
momentum distribution in Table 1, we know that
the asymptotic slope parameter $T_s$ has possible
origins different from each other for the above two models:
The critical temperature dominates the asymptotic slope parameter
in the QGP fluid model with phase transition,
while the initial
temperature dominates the asymptotic slope parameter
in the hot hadron gas model without phase transition.

\section{CONCLUDING REMARKS}
We have derived the thermal photon distribution emitted
from a hot matter produced by the high energy nuclear collisions, 
based on hydrodynamical model, and compared these theoretical results 
with S+Au 200 GeV/nucleon data obtained by CERN WA80. 
We have observed that only the QGP fluid (with phase
transition) model can consistently reproduce the above 
experimental data.
Furthermore we have determined the asymptotic slope 
parameter of the transverse photon distribution through
comparison of our numerical results of hydrodynamical
simulation with the thermal photon distribution. 

\begin{center}
\bf ACKNOWLEDGMENT
\end{center}
\rm 

The authors are much indebted Professor I.\,Ohba for his 
helpful comments.
They also thank H.\,Nakamura and other members of 
high energy physics group of Waseda Univ.\ for their discussion.
Numerical calculation has been done with work-stations of
Waseda Univ.\ high energy physics group.

\appendix
\begin{center}
\bf APPENDIX : QUANTUM LANGEVIN EQUATION 
\end{center}
\rm 
We introduce a quantum Langevin equation for the 
annihilation operator
\BE \label{Q-L_eq}
i\frac{d}{dt}a({\V p},t)=\int^{t}K({\V p},t')a({\V p},t')dt'
+f({\V p},t),
\EE
and its hermit conjugate, where $K({\V p},t)$ and $f({\V
p},t)$ are input parameter and random force, respectively.
Canonical operators $a(\V{p},t)$ and $a^{\dagger}(\V{p},t)$ should
satisfy following two requirements:

1) Equal time commutation relation
\BE
[a(\V{p},t),a^{\dagger}(\V{p},t)]_{\mp} =\delta^3(\V{p}-\V{p}'),
\EE
where  $\mp$ stand for boson and fermion respectively.

2) The Kubo-Martin-Schwinger condition
\BE \label{KMS}
\langle a^{\dagger}(\V{p},t)a(\V{p}',t'+i\beta)\rangle
=\langle a(\V{p}',t')a^{\dagger}(\V{p},t) \rangle
{\rm e}^{\beta\mu}.
\EE

We can fix the detailed properties of $a$ and $f$ so as to
satisfy above requirements.

The random force operator can be expanded as
\BE \label{random}
f(\V{p},t)=\int_{-\infty}^{+\infty}d\omega'
\bigl[\frac{\gamma(\V{p},\omega')}{2\pi}\bigr]^{1/2}
\rho(\V{p},\omega')A(\V{p},\omega')
{\rm e}^{-i\omega't},
\EE
where $A(\V{p},\omega)$ is a {\it canonical }random operator whose 
canonical commutation relation is given by
\BE
\bigl[A(\V{p},\omega),A^{\dag}(\V{p}',\omega')\bigr]
=\delta^3(\V{p}-\V{p}')\delta(\omega-\omega').
\EE
The stationary solution of Eq. (\ref{Q-L_eq}) is given as
\BE
a(\V{p},t)=\int_{-\infty}^{\infty}d\omega
\sqrt{\frac{\gamma(\V{p},\omega)}{2\pi}}
\frac{\rho(\V{p},\omega)A(\V{p},\omega)}
{\omega-E(\V{p},\omega)}{\rm e}^{-i\omega t}.
\EE
The Kubo-Martin-Schwinger
condition (\ref{KMS}) can be satisfied by the following ansatz:
\BE \label{correlation}
\begin{array}{lcl}
\langle A(\V{p},\omega)\rangle 
&=& \langle A^{\dag}(\V{p},\omega)\rangle = 0,
\\
\langle A^{\dag}(\V{p},\omega)A(\V{p}',\omega')\rangle
&=& \delta^3(\V{p}-\V{p}')\delta(\omega-\omega')n(\omega,T),
\\
\langle A(\V{p},\omega)A^{\dag}(\V{p}',\omega')\rangle
&=& \delta^3(\V{p}-\V{p}')\delta(\omega-\omega')
\bigl[1+\xi n(\omega,T)\bigr],
\\
\langle A(\V{p},\omega)A(\V{p},\omega) \rangle
&=& \langle A^{\dag}(\V{p},\omega)A^{\dag}(\V{p},\omega) \rangle=0,
\end{array}\label{ave}
\EE
\BE
n(\omega,T)=\frac{1}{\exp \bigl(\frac{\omega-\mu}{T} \bigr)
-\xi } \ , \nonumber 
\EE
where $\xi$ takes $+1$ for boson and $-1$ for fermion. 

From (\ref{random}) and (\ref{correlation}), {\it quantum}
fluctuation-dissipation theorem for the operator-valued 
Langevin equation (\ref{Q-L_eq}) is obtained
\BE
\begin{array}{ccl}
\langle f(\V{k},t) \rangle &=&\langle f^{\dag}(\V{p},t) \rangle =0
\\
\langle f^{\dag}(\V{p},t)f(\V{p}',t')\rangle
&=& \delta^3(\V{p}-\V{p}')\displaystyle{\int_0 ^{\infty}}
\displaystyle{\frac{d \omega}{2 \pi}}
\rho^2(\V{p},\omega)\gamma(\V{p},\omega)
n(\omega,T){\rm e}^{-i\omega(t-t')}
\\
\langle f(\V{p},t)f^{\dag}(\V{p}',t')\rangle
&=& \delta^3(\V{p}-\V{p}')\displaystyle{\int_0 ^{\infty}}
\displaystyle{\frac{d \omega}{2 \pi}}
\rho^2(\V{p},\omega)\gamma(\V{p},\omega)
\bigl[1+\xi n(\omega,T)\bigr]{\rm e}^{+i\omega(t-t')}
\\
\langle f(\V{p},t)f(\V{p}',t') \rangle &=&
\langle f^{\dag}(\V{p},t)f^{\dag}(\V{p}',t') \rangle = 0
\end{array}
\EE

We have formulated a quantum Langevin equation consistently.
We make thermal field operators for bosons (\ref{b-field})
and fermions (\ref{f-field}) by means of the above operator
$a(\V{p},t)$ and $a^{\dag}(\V{p},t)$.


\newpage

\begin{center}
Table 1\\
\begin{tabular}{c|c|c|c|c|c}
\hline
Model & $T$ (MeV)& $v_T$ & $v_L$  & $T_{eff}$  (MeV) & $T_s$ (MeV)\\
\hline
QGP fluid (QGP phase) & 157.5 & 0.53 & 0.11 & 280.8 &
273.2\\
QGP fluid (hadron phase) & 157.5 & 0.53 & 0.11 & 280.8 & 
270.5\\
\hline
Hot hadron gas & 400.0 & 0.0 & 0.0 & 400.0 & 390.1\\
\hline
\end{tabular}
\end{center}

\begin{large}

TABLE CAPTION

\end{large}

The maximum $T_{eff}$ for the QGP fluid model and the hot
hadron gas model in our hydrodynamical simulation, and the slope
parameters $T_s$ at $k_T$ = 20 GeV in Fig.\,8.
$T_{eff}$ is evaluated by Eq.\,(26).

\begin{Large}

FIGURE CAPTION

\end{Large}

FIG.\,1  The phase transition-like behavior of energy density
distribution. The solid curve stands for our QGP fluid model 
with phase transition, 
the dotted curve for the fluid of the hot hadron gas composed 
of massive pions and kaons, and the dashed curve for the fluid of 
the quark-gluon plasma composed of massless u-, d- s-quarks
and gluons. The critical temperature $T_c=160$ MeV.

FIG.\,2 The pseudo-rapidity distribution of charged
hadrons in S+Au 200 GeV/nucleon collision. The experimental data
was obtained by CERN WA80[12]. The solid curve and the dashed
curve  stand for, respectively, the QGP fluid model
with phase transition and the hot hadron gas model without
phase transition.

FIG.\,3 The transverse momentum distribution of neutral
pions in S+Au 200 GeV/nucleon collision. The experimental data was
obtained by CERN WA80[13]. The solid curve and the dashed
curve stand for, respectively, the phase transition
model and the hot hadron gas model.

FIG.\,4 Three kinds of photon emission processes represented by
Eqs.\,(19) and (20). The double line stands for
off-shell particles in the heat bath. The wavy line stands
for the external line of the photon.

FIG.\,5 The production rate as a function of temperature
represented by Eq.\,(18). The solid curve stands for the phase
transition model, the dotted curve for the production rate of 
the hadron phase and the dashed curve for the production rate
of quark-gluon plasma phase. The critical temperature $T_c=160$
MeV.

FIG.\,6 The transverse momentum distribution of thermal photon
in S+Au 200 GeV/nucleon collision. The upper limit of
experimental data was estimated by CERN WA80[15]. The solid
curve and the dotted curve are, respectively, the
contribution of the hadron phase region and the QGP phase
region in the QGP fluid model
with phase transition. The dashed curve is the photon
distribution of the hot hadron gas model without phase
transition.

FIG.\,7 The transverse momentum distribution at rapidity
$y=0$. The solid curve and the
dotted curve are, respectively, the contribution of the
hadron phase region and the QGP phase region in the phase
transition model. The dashed curve is the photon
distribution of the hot hadron gas model. 

FIG.\,8 The slope parameter of the transverse momentum
distribution which is shown in Fig.\,7. 
The solid curve and the
dotted curve stand for, respectively, the
hadron phase and the QGP phase in the QGP fluid model
with phase transition. The dashed curve is the photon
distribution of the hot hadron gas model without phase transition.
The solid bold line and the dashed bold line stand for,
respectively, $T_{eff}$ of the QGP fluid model and $T_{eff}$ of
the hot hadron gas model.
The values of the asymptotic slope parameter $T_{eff}$ are
evaluated by Eq.\,(26) and shown in Table 1.

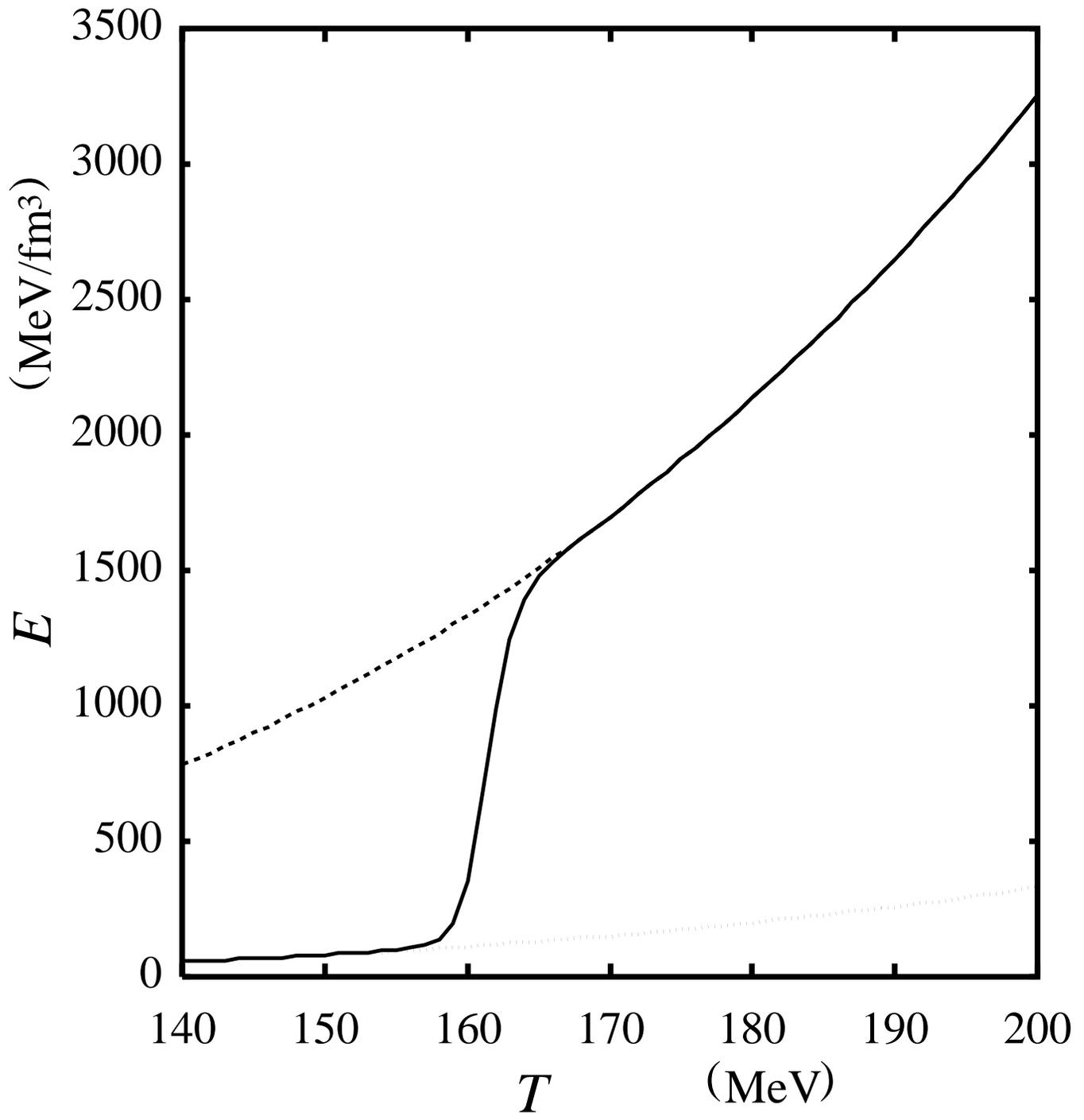
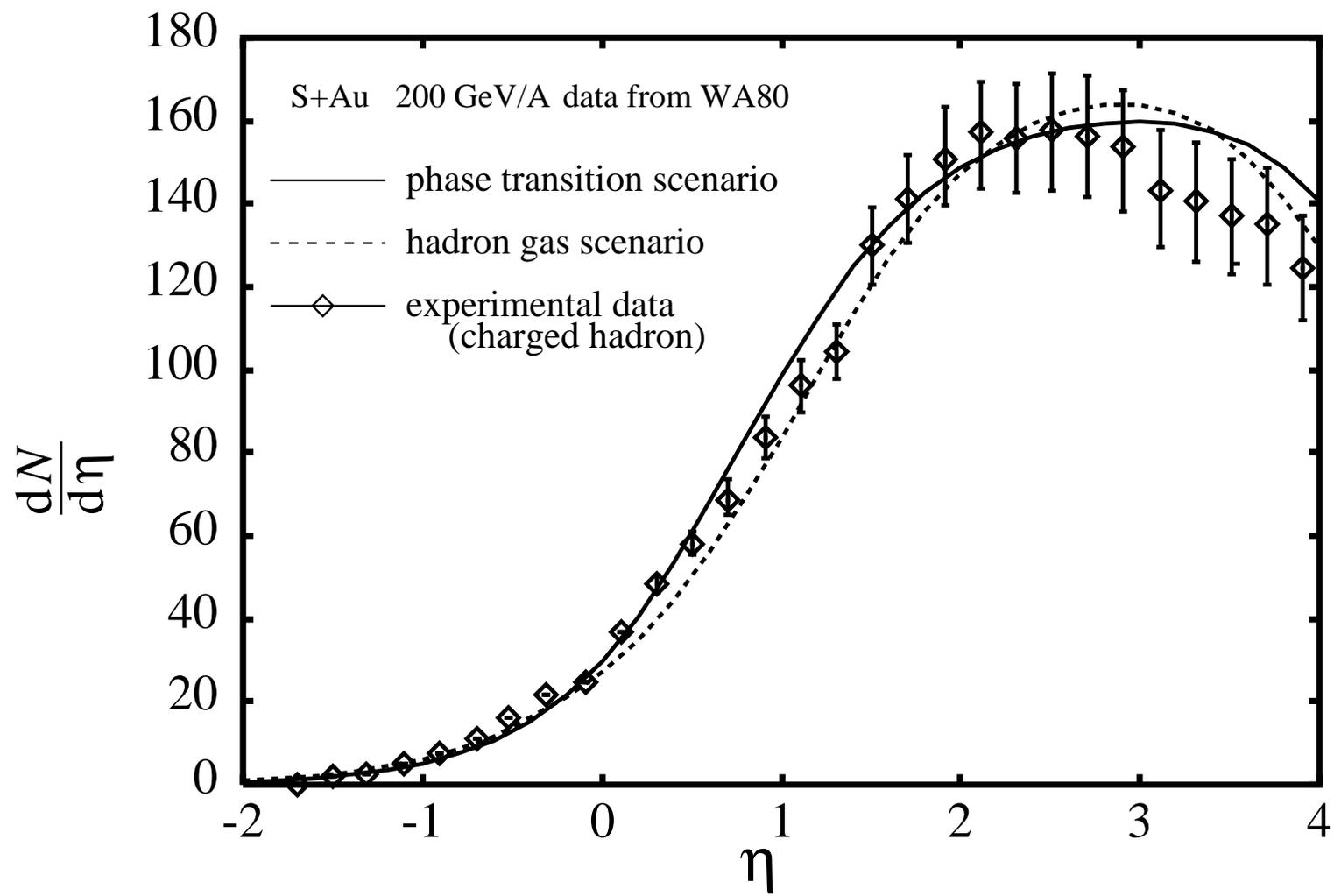
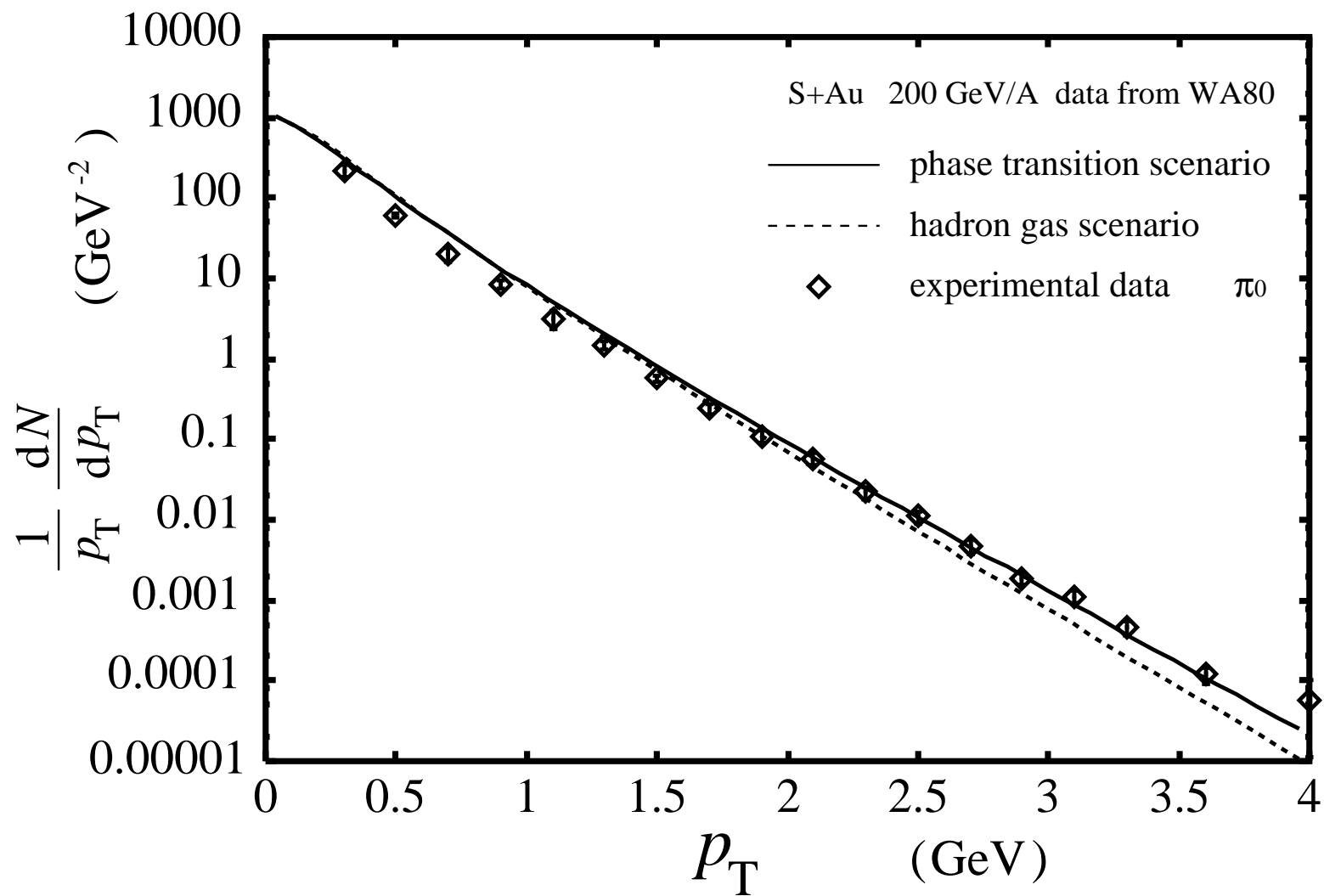
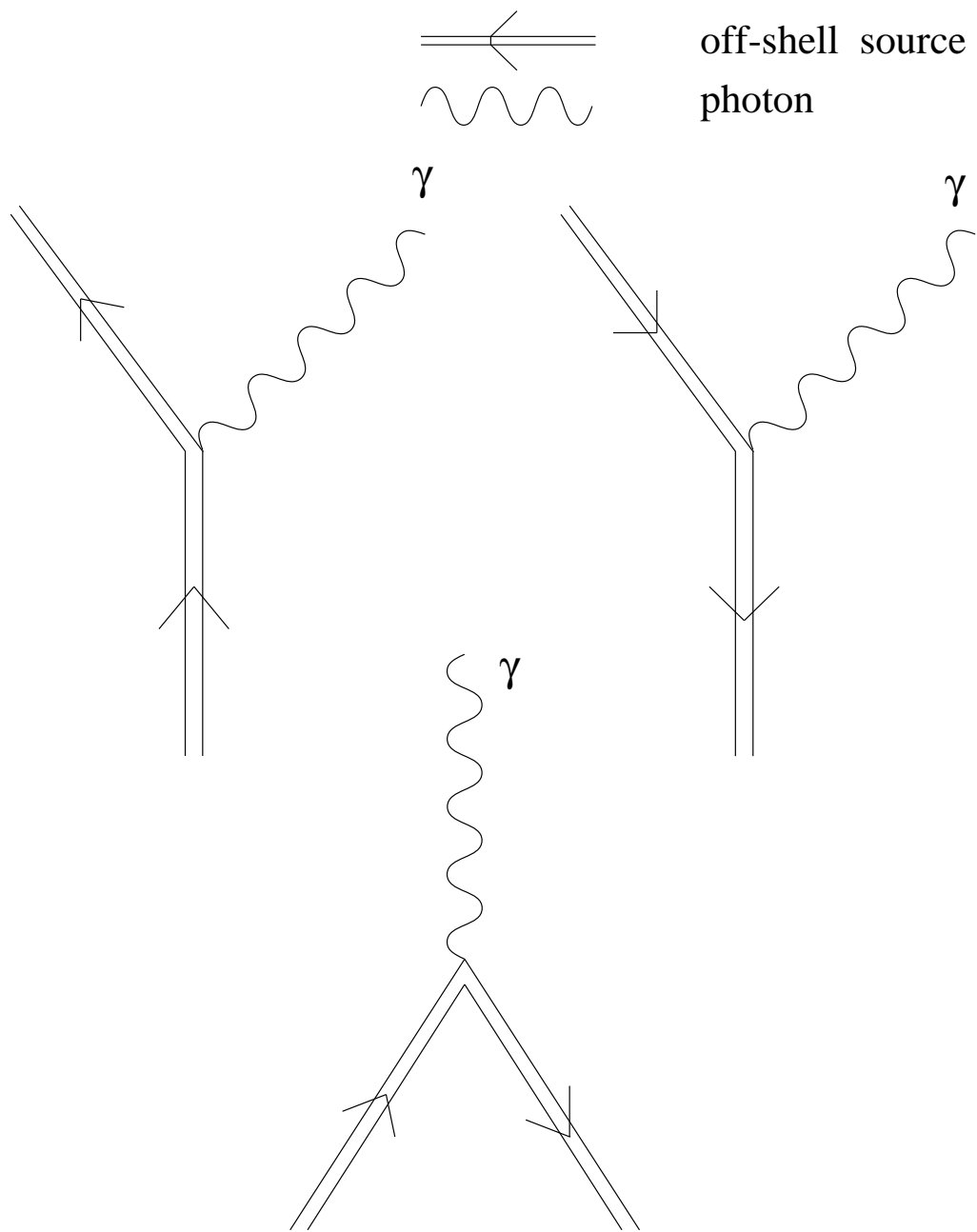
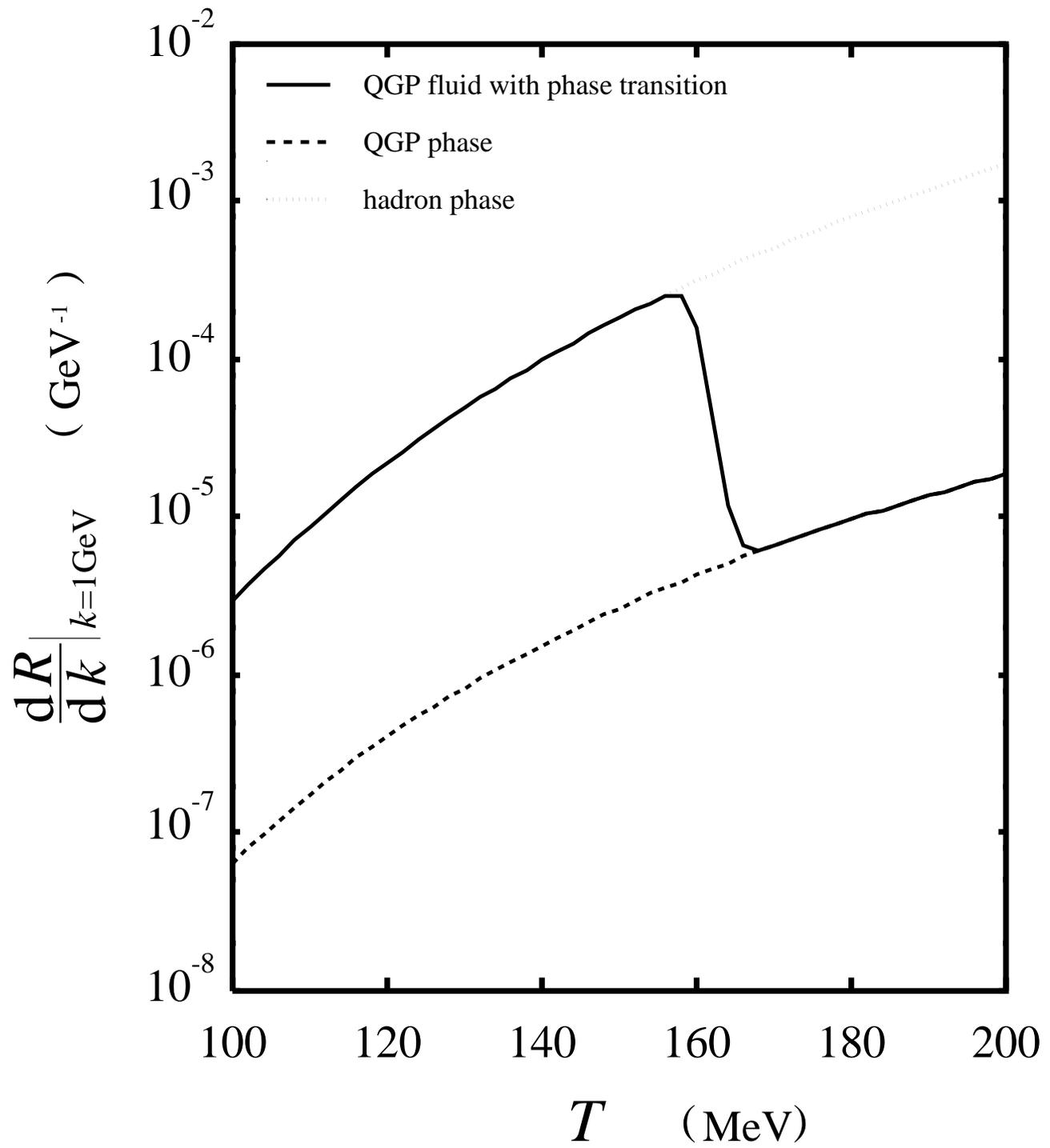
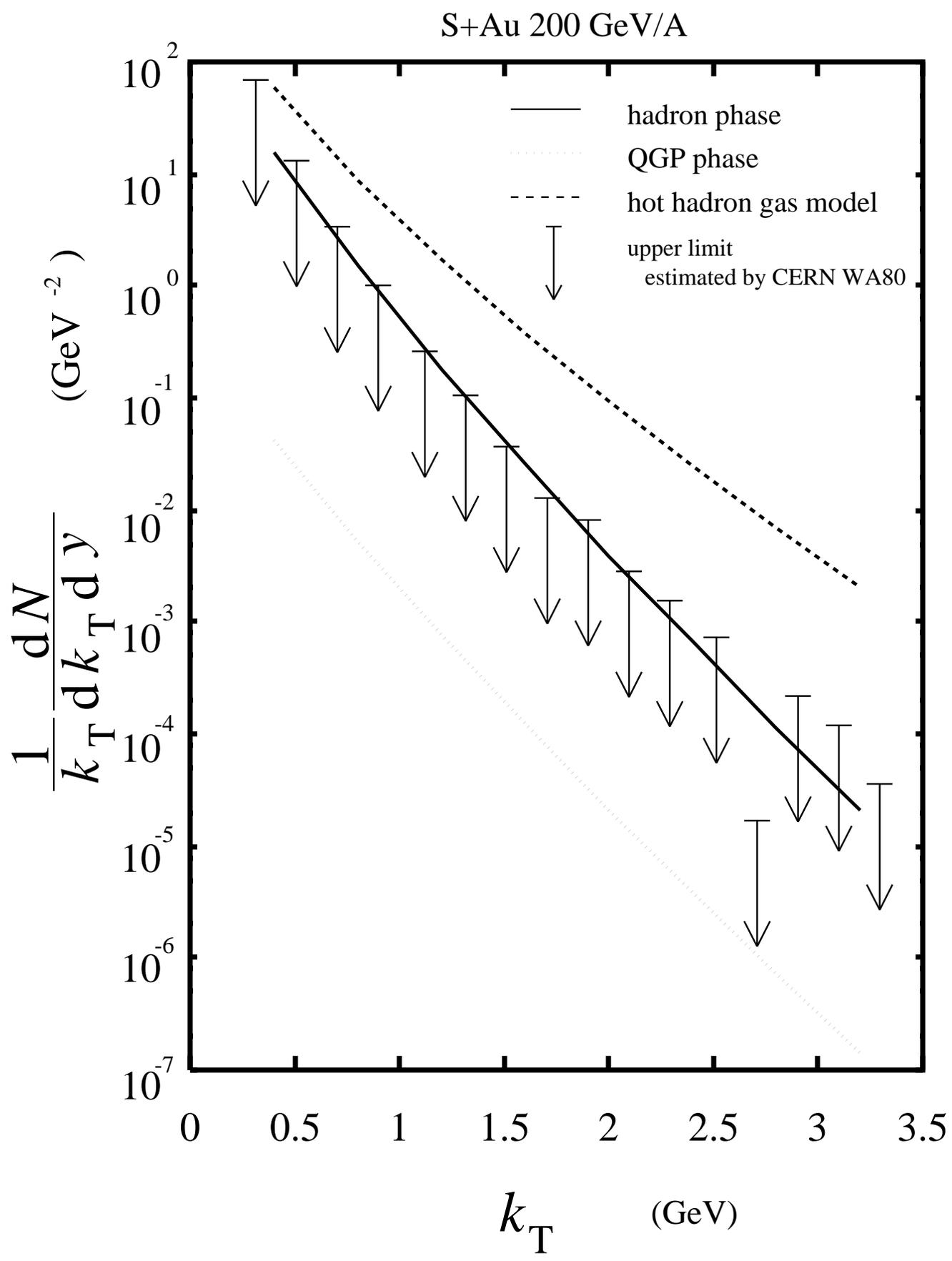
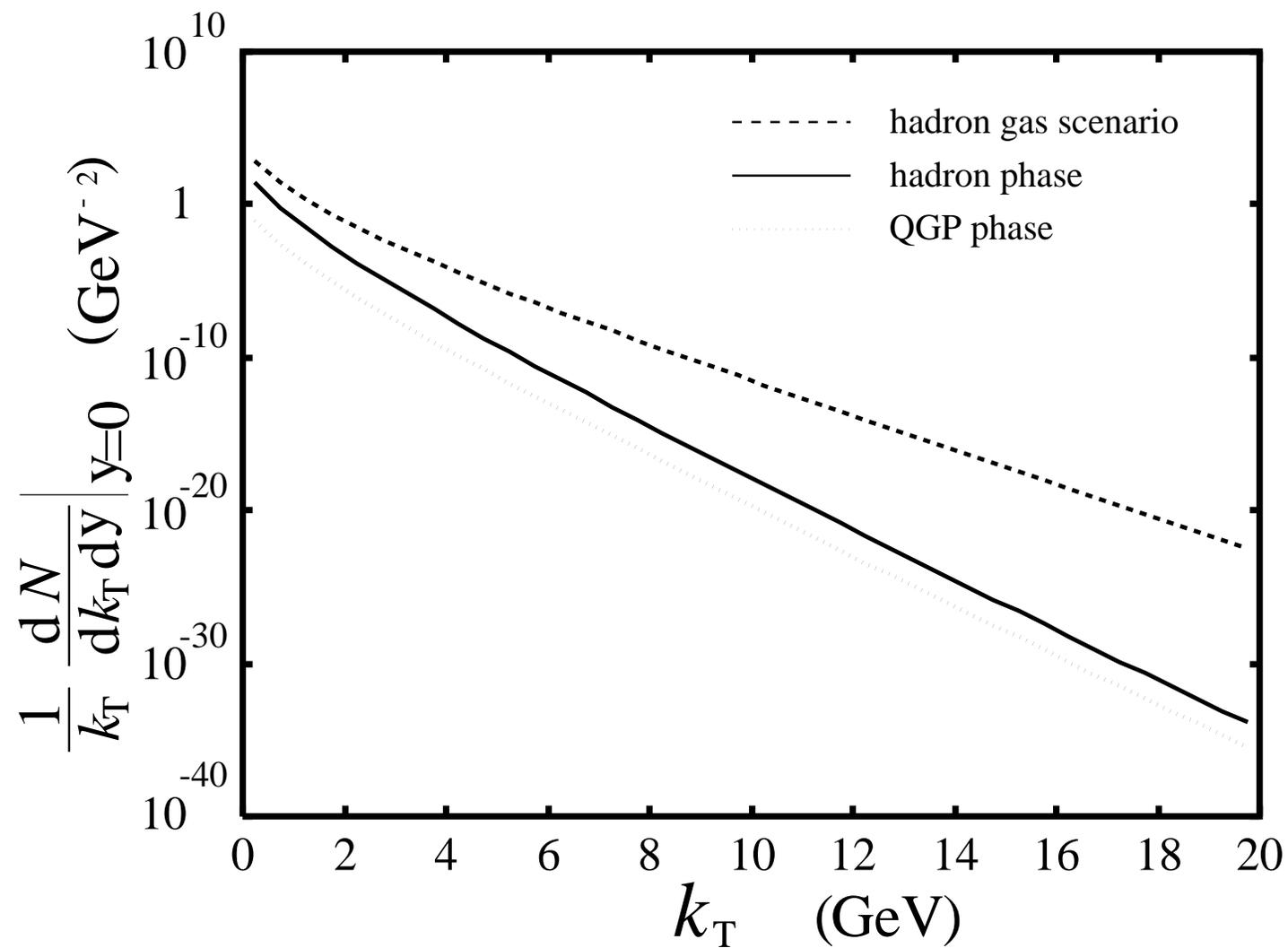
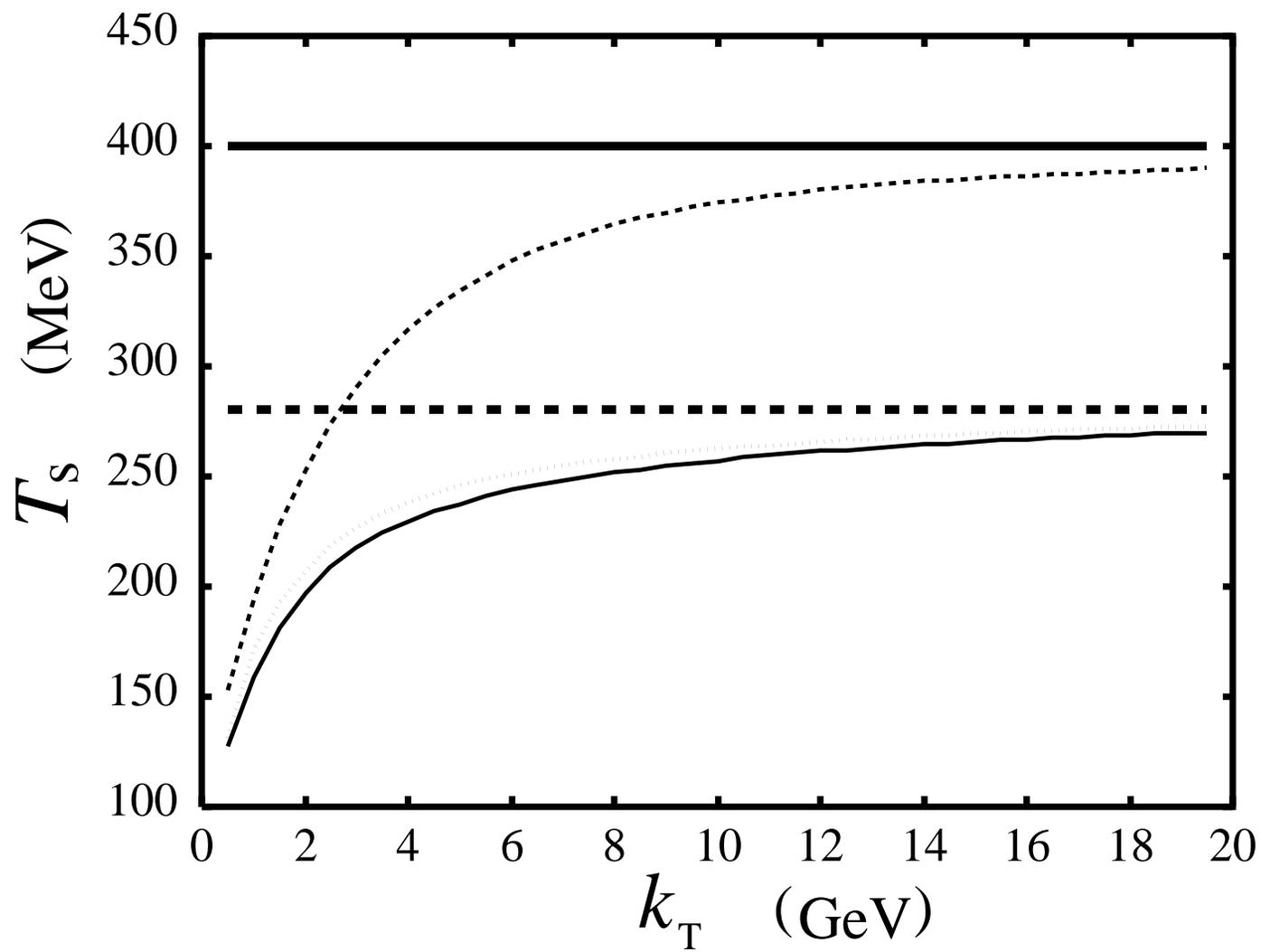
\end{document}